%% file: tensor26.tex
\documentclass[final,5p,times,twocolumn]{elsarticle}
 \biboptions{comma,sort&compress}
\usepackage{graphicx}
\usepackage{amsmath}
\usepackage{here}
\usepackage[dvips]{epsfig}

\usepackage{xcolor}
\usepackage{soul}
\usepackage{ulem}
\usepackage{hyperref}
\newcommand{\ta}[1]{\textcolor{blue}{#1} }

\def\beq{\begin{equation}}
\def\eeq{\end{equation}}
\def\bea{\begin{eqnarray}}
\def\eea{\end{eqnarray}}
\def\bq{\begin{quote}}
\def\eq{\end{quote}}

\def\nnb{\nonumber}
\def\ga{\left(}
\def\dr{\right)}

\def\lrar{\longrightarrow}

\def\nnb{\nonumber}
\def\la{\langle}
\def\ra{\rangle}
\def\nin{\noindent}
\def\ba{\vspace*{-0.2cm}\begin{array}}
\def\ea{\end{array}\vspace*{-0.2cm}}

\def\d{$\diamond~$}
\def\als{\alpha_s}

\def\gg2{ \la\alpha_s G^2 \ra}
\def\gg3{g^3f_{abc}\la G^aG^bG^c \ra}
\def\ggg4{\la\als^2G^4\ra}

\journal{Elsevier}

\begin{document}

\begin{frontmatter}

\title{ $2^{++}$ Light Tensor  Hybrid Meson from QCD Laplace Sum Rules}
 
\author[label0]{Jason Ho}
\address[label0]{Department of Physics, Dordt University, Sioux Center, Iowa, 51250, USA}
\ead{jason.ho@dordt.edu}

\author[label1]{Robin Kleiv}
\address[label1]{Department of Physics, Thompson Rivers University, Kamloops, BC, V2C 0C8, Canada}
\ead{rkleiv@tru.ca}

 \author[label2]{Siyuan Li}
\address[label2]{Institut f\"ur Kernphysik,
 Johannes Gutenberg-Universit\"at  Mainz,  D-55128 Mainz, Germany}
   \ead{siyuan.li@uni-mainz.de} 
   
 \author[label3,label4]{Stephan Narison}
\address[label3]{Laboratoire
Univers et Particules de Montpellier, CNRS-IN2P3, 
Case 070, Place Eug\`ene
Bataillon, 34095 - Montpellier, France}
\address[label4]{Insitute of High-Energy Physics (iHEPMAD), Univ. Ankatso, Antananarivo, Madagascar}
  \ead{snarison@yahoo.fr}  
  
   \author[label5]{Tom  Steele}
   \address[label5]{Department of Physics and  Engineering Physics, University of Saskatchewan, SK, S7N 5E2, Canada}
   \ead{tom.steele@usask.ca}
   
    \author[label4]{Davidson Rabetiarivony}
   \ead{rd.bidds@gmail.com}

\begin{abstract}
\nin
We present an analysis of the light tensor  ($J^{PC}=2^{++}$) hybrid 
 {meson mass and coupling} from QCD Laplace Sum Rules where the next-to-leading order (NLO) perturbative (PT) corrections and the contributions of the non-perturbative (NP) condensates up to dimension-six ($D=6$) are included. NLO leading-logarithms corrections due to the condensates which contribute in the chiral limit are considered. 
We obtain the mass {$M_{2^+}= (2038\pm 190)$} MeV and a relatively small coupling $f_{2^+}={(10.5\pm 2.9)}$ MeV normalized as $f_\pi=93$ MeV. Our results suggest that the $f_2(1950)$ or/and the $f'_2(2010)$ may have a sizeable $\bar qqg$ hybrid component. We also compute the tensor hybrid topological charge (value of the two-point  function at zero momentum) and find (for the first time) at NLO\,: {$\Pi_{qg}(0)=\ga 2.41\pm 0.43\dr \times 10^{-4}\,{\rm GeV}^6$} which could be checked from {some   lattice QCD  or/and low energy theorems} (LET).

\end{abstract}
\end{frontmatter}
\section{Introduction}
The analysis of the light tensor meson channel can help to understand the nature of the different experimental candidates quoted by the PDG\,\cite{PDG}. Different works on the tensor mesons have been reported in the past, including the following:

\d The spectrum of the $2^{++}$ gluonium has been studied  in the pioneering work of Novikov et al. (NSVZ)\,\cite{NSVZ} and studied using QCD spectral sum rules (QSSR)\,\cite{SVZ}\,\footnote{For reviews, see e.g. \cite{SNB2,SNB1,SNREV22,SNB26,DERAF,RRY}.} in\,\cite{SNG0,SNB2,SNG} and more recently in \cite{SLI} at NLO where the gluonium mass and the Renormalization Group Invariant (RGI) coupling are\,:
\beq
M_{2g}=3028(287)~{\rm  MeV},~~~~~~ \hat f_{2g}=173(23)~{\rm MeV}. 
\eeq
where--here and in the following--the coupling is normalized as $f_\pi=93$ MeV. 
The inclusion of the NLO correction has shifted the leading-order (LO) mass of about 2 GeV\, \cite{SNG} to a higher value which does not favour some eventual gluonia/glueball interpretation of the observed  $f_2(2010,2300,2340)$ {states\,\cite{KLEMPT}}.   

\d The spectrum of the $\bar qq$ $2^{++}$ tensor meson has also been studied at NLO within QSSR in Ref.\,\cite{BAGANT}\ where the RGI coupling was safely determined but not the mass:
\beq
M_{f_2q}= (1414\sim 1580)~{\rm MeV} ,~~~\hat f_{f_2q}= (183\sim 284)~{\rm MeV}. 
\eeq

\d $\bar qq$ and gluonium tensor mesons mixing has been studied in Ref.\,\cite{BRAMON} using an off-diagonal two-point function where a small mixing angle of about 10$^\circ$ has been found. 

\d In this paper, we complete the analysis by studying the mass and coupling of the light tensor ($J^{PC}=2^{++}$) hybrid meson using QSSR at NLO of PT and including contributions of the QCD NP condensates up to dimension-six.  NLO leading-logarithms corrections due to the condensates which contribute in the chiral  limit are also included. 
\section{The $2^+$ light hybrid QCD two-point function}

We shall be concerned with the spin-2 component of the two-point function\,\footnote{For relations among different form factors,  see e.g.\,\cite{CNZa}.}:
\bea
&\hspace*{-0.5cm} \psi_{qg}^{\mu\nu\rho\sigma}(q^2)\equiv i\int d^4x \,e^{iqx}\la 0 \vert J^{\mu\nu}_{qg}(x) \ga J^{\rho\sigma}_{qg}\dr^\dagger(0)\vert 0\ra\nnb\\
&=\ga P^{\mu\nu\rho\sigma}\equiv \tilde\eta^{\mu\rho}\tilde\eta^{\nu\sigma}+\tilde\eta^{\mu\sigma}\tilde\eta^{\nu\rho}-\frac{2}{n-1}\tilde\eta^{\mu\nu}\tilde\eta^{\rho\sigma}\dr \psi_{qg}(q^2),
\label{eq:2-point}
\eea
built from the hybrid tensor charged current\,:\,\footnote{Some other choices of currents can be found in\,\cite{TAN}. We shall comment this work in the conclusion. } :
\beq
\hspace*{-0.5cm}J^{\mu\nu}_{qg}(x)=g_s \bar u_i^\gamma\sigma^{\mu\alpha}_{ij} G^a_{\alpha\nu} \frac{\lambda^a_{\gamma\delta}}{2} d_j^\delta~, 
\label{eq:current}
\eeq
with
\beq
\tilde\eta^{\mu\nu} \equiv q^2g^{\mu\nu}-q^\mu q^\nu \,,\quad P_{\mu\nu\rho\sigma} P^{\mu\nu\rho\sigma}=2(n^2-n-2),
\label{eq:proj}
\eeq
where $g_s$ is the QCD coupling; $u,d$ are the up and down light quark fields; $G^a_{\mu\nu}$ is the gluon field strength; $\lambda^a$ is the colour matrix; $\sigma_{\mu\nu}  \equiv \frac{i}{2}[\gamma_\mu,\gamma_\nu]$ where $\gamma_{\mu}, \gamma_{\nu}$ are the Dirac matrices; and
$n=4+2\epsilon$ is the space-time dimension used for dimensional regularization and renormalization in the $\overline{\rm MS}$ scheme.
\section{The two-point function at LO up to $D=6$ condensates}
\d To LO and up to $D=6$ quark and gluon condensates, the QCD expression is  (see the details of each contribution in Fig.\,\ref{fig:ope})\,:
\begin{figure}[hbt]
\vspace*{-0.5cm}
\begin{center}
\includegraphics[width=11cm]{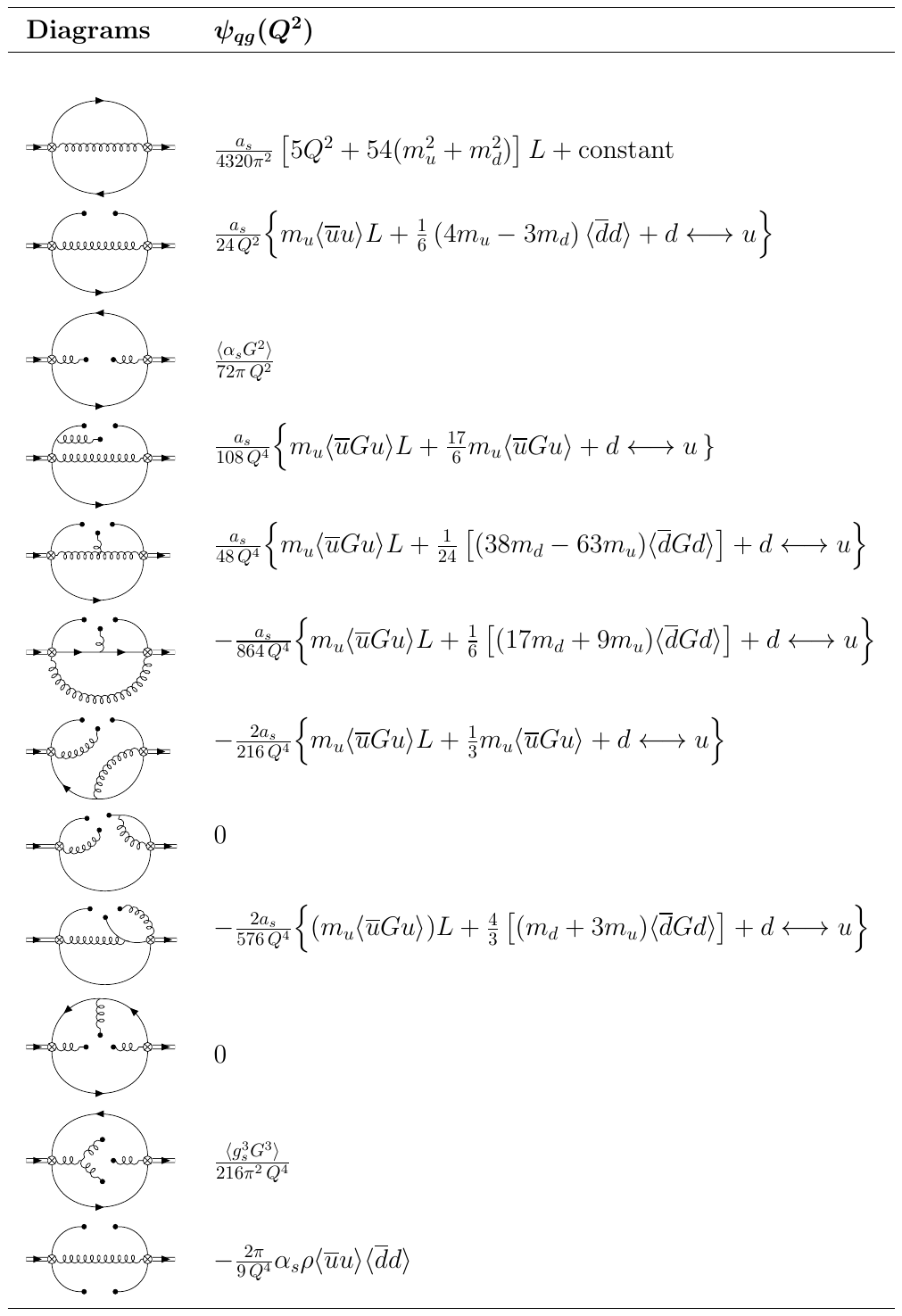}
\vspace*{-1.5cm}
\caption{\footnotesize  QCD expression of the two-point correlator at LO of PT series and including dimension-six condensates contributions. Topologically equivalent permutations of Feynman diagrams are not shown. $\rho$ indicates the deviation from four-quark factorization.}
\label{fig:ope}
\end{center}
\vspace*{-0.8cm}
\end{figure} 
\bea
\psi_{qg}\vert_{LO}(Q^2)&=& Q^2\Bigg{\{} \Big{[}d_0 +\frac{d_2}{Q^2}  +\frac{d_{4q}}{Q^4} +
\frac{d_{6qg}}{Q^6}\Big{]}L+\nnb\\ 
&& \frac{d_{4g}+d^c_{4q}}{Q^4} + \frac{d_{6q}+d_{6g}+d^c_{6qg}}{Q^6} +\cdots \Bigg{\}}
\label{eq:ope}
\eea
where $Q^2\equiv -q^2$, $L\equiv \log {(Q^2}/{\nu^2)}$. The coefficients $d_{2np}$ are  the contributions of the condensates $p\equiv$ gluon, quark or quark-gluon fields  of dimension $2n$ entering in the Operator Product Expansion (OPE). Their expressions can be deduced from Fig.\,\ref{fig:ope} where the sum of the mixed condensate contributions{, including  topological multiplicity factors of 2 for diagrams 7--9,} is 
\bea
d_{6qg}&=&\frac{7a_s}{432}\ga m_u\la\bar u G u\ra + m_{d}\la\bar d 
Gd\ra\dr ,\nnb\\
d^c_{6qg}&=&\frac{a_s}{32}\Bigg{[}\frac{125}{81} \ga m_u\la\bar u G u\ra + m_{d}\la\bar d
Gd\ra\dr \nnb\\
&&- \frac{9}{4}\ga m_u\la\bar d G d\ra + m_{d}\la\bar u
Gu\ra \dr\Bigg{]}.
\label{eq:ope_terms_lo}
\eea
We denote\: $a_s\equiv \alpha_s/\pi$ and
\beq 
\hspace*{-0.5cm}
\la\bar \psi G \psi\ra \equiv \la g_s \bar \psi_i^\alpha\sigma^{\mu\nu}_{ij} G^a_{\mu\nu} \frac{\lambda^a_{\alpha\beta}}{2} \psi_j^\beta\ra =M_0^2\la\bar \psi\psi\ra~:~~~\psi \equiv u,d.
\eeq
 We shall use the input parameters given in Table\,\ref{tab:param}.

\d To complete the analysis, we also computed the value of the two-point function
at $Q^2=0$ and found
\beq
\psi_{qg}\vert_{LO}(Q^2=0) \sim  m^2_i \log{\frac{ m_i^2}{\nu^2}}
\eeq
indicating that the two-point function has smooth LO behaviour at $Q^2=0$. One may parameterize its NP part by the ground state contribution using a dispersion relation  proportional to
$f_{2^+}^2$ where the coupling is normalized as in Eq.\,\ref{eq:mda}.

\section{The lowest Laplace sum rule moments}
\d We shall be concerned with  the following inverse Laplace transform moments and their ratio\,\,\cite{SVZ,SNR,SN23,BELLa,BERTa}\,:
\bea
{\cal L}_{0,1}^c(\tau,\nu)
&=&\int_{t>}^{t_c}dt~t^{(0,1)} e^{-t\tau}\frac{1}{\pi} \mbox{Im}\,\psi_{qg}(t,\nu)~,
\nnb\\
 {\cal R}^c_{10}(\tau)&\equiv&\frac{{\cal L}^c_{1}} {{\cal L}^c_0}= \frac{\int_{t>}^{t_c}dt~e^{-t\tau}t\, \mbox{Im}\,\psi_{qg}(t,\nu)}{\int_{t>}^{t_c}dt~e^{-t\tau} \mbox{Im}\,\psi_{qg}(t,\nu)}.
\label{eq:lsr}
\eea

\d To get the Finite Energy Sum Rule (FESR) lowest moment ${\cal L}_{0}^c$, we take the second $Q^2$ derivative of the two-point function,  which is superconvergent and thus obeys an homogeneous Renormalization Group Equation (RGE). Then, we parameterize the spectral function by the Minimal Duality Ansatz (MDA)\,:
\beq
\hspace*{-0.25cm}\frac{1}{\pi}{\rm Im}\,\psi_{qg}(t)= 2f_{2^+}^2 M_{2^+}^2\delta (t-M_{2^+}^2)+\theta \ga t-t_c\dr``QCD~continuum",
\label{eq:mda}
\eeq
where we assume that the QCD expression of the spectral function above the continuum threshold $t_c$ smears all radial excitation contributions. The coupling $f_{2^+}$ is normalized as $f_\pi=93$\,MeV. The QCD-continuum  contribution is quantified by:
\beq
\rho_n=e^{-t_c\tau}\ga 1+(t_c\tau)+\cdots +\frac{(t_c\tau)^n}{n !}\dr~,
\eeq
which we transfer to the QCD side of the sum rule.

\d The second ${\cal L}_{1}^c$ moment is obtained by taking the $\tau$-derivative of ${\cal L}_{0}$, transferring the QCD-continuum  contribution to the {QCD side} of the sum rule.  

\d  In the MDA parameterization, the ratio of moments
is equal to the mass squared of the ground state
\beq
{\cal R}_{10}^c\simeq  M_h^2,
\eeq
while the lowest moment can provide a prediction of the ground state coupling constant $f_h$ once the value of the mass is known.

\section{QCD expression of the Laplace sum rule at LO}
Taking the Laplace transform of the second derivative of the two-point function by using the master Laplace sum rule formulae given in the Appendices of Refs.\,\cite{SNB2,SNB1}:
\bea
\hspace*{-0.75cm}
 {\cal L}\Big{[}\frac{1}{(Q^2+t)^{\alpha+1}}\Big{]}&=& \frac{\tau^{\alpha+1}}{\Gamma(\alpha+1)}  e^{-t\tau}\nnb\\
{\cal L}\Bigg{[} \frac{ \log\ga\frac{Q^2}{\nu^2}\dr}{(Q^2)^{\alpha+1}}\Bigg{]}&=& \frac{ \tau^{\alpha+1}}{\Gamma(\alpha+1)}  \big{[}-\log \tau\nu^2 +\psi(\alpha+1)\big{]}.
\eea
 $\psi(z)=-\frac{d}{dz} \log \Gamma(z) $ is the Digamma or $\psi$ function with $\psi(1)=-\gamma_E =-0.5772...$,  the Euler-Mascheroni constant. Then, we deduce the Laplace transform for $\tau=1/\nu^2$ 
{(see e.g., Ref.~\cite{SNR} for discussion of RGE scale in Laplace sum-rules) }
 at LO of the PT series\,:
\bea
\hspace*{-0.5cm}
{\cal L}_{0}^c(\tau)\vert_{\rm LO}&=&\tau^{-2}\Bigg{\{}d_0(1-\rho_2) -d_2\tau-(d_{4q}\gamma_E-d_{4q}^c-d_{4g})\tau^2\nnb\\
&+&\Big{[}(d_{6g}+d_{6q}+d^c_{6qg})+d_{6qg}(1-\gamma_E)\Big{]}\tau^3\Bigg{\}},
\label{eq:srlo}
\eea
 from which, one can derive ${\cal L}_{1}^c$ and thus
${\cal R}_{10}^c$.
\section{QCD input parameters}
We shall use the most recent phenomenological determinations from QCD spectral sum rules,
$e^+e^-\to$ Hadrons and $\tau$-decay data {as shown in Table~\ref{tab:param}}. 

\begin{table}[H]
\vspace*{-0.5cm}
\setlength{\tabcolsep}{1.2pc}
    {\small
\begin{tabular}{lll}
&\\
\hline
 Parameters& Values& Refs.    \\
\hline
$\alpha_s(M_\tau)$& $0.3128(58)$&\cite{SNREV25}\\
$\hat{m}_u$&$2.9(3)$ MeV& \cite{SNlmass}\\
$\hat{m}_d$&$5.7(6)$ MeV & \cite{SNlmass}\\
$\la\alpha_s G^2\ra$& $(6.35\pm 0.35)\times 10^{-2}$ GeV$^4$&
\cite{SNG2}\\
$M_0^2$&$(0.8 \pm 0.2)$ GeV$^2$&\cite{JAMI2,HEID,SNhl}\\
$\la g^3  G^3\ra$& $(8.2\pm 1.0)$ GeV$^2\times\la\alpha_s G^2\ra$&
\cite{SNG3}\\
$\rho \alpha_s\la \bar qq\ra^2$&$(6.38\pm 0.30)\times 10^{-4}$ GeV$^6$&\cite{SN4q}\\
\hline

\end{tabular}
}
\vspace*{-0.25cm}
    \caption{QCD input parameters. } 
    \label{tab:param}
\end{table}

The value of $\alpha_s$ in Table\,\ref{tab:param} corresponds to a QCD scale of
\beq
\Lambda = (324\pm 10)~{\rm MeV}~~{\rm for~ 3~ flavours},
\eeq
evaluated at three-loops accuracy\,\footnote{In the following, we shall use this value and (for a consistency) the three-loops expression of $\alpha_s$.}. 
The invariant masses $\hat{m}_u\,, \hat{m}_d$ are related to the running masses evaluated at {$\nu=1/\sqrt{\tau}$}\,\cite{FNR,SNB1,SNB2}\,:
\beq
\bar m_i(\tau) =\frac{\hat m_i}{{\ga-\log {\tau}{\Lambda^2}\dr}^{\frac{\gamma_1^m}{-\beta_1}}}\Big{[} 1+\frac{145}{162}a_s(\tau)+{\cal O}(a_s^2)\Big{]},
\eeq
where ${\gamma_1^m} =2$ and $-\beta_1=(1/2) (11 -2n_f/3)$ are the LO mass anomalous dimension and $\beta$ function for $n_f$ flavours while the $\alpha_s$ correction is given numerically for three flavours. 
We estimate the product $m_q\la\bar qq\ra$ using the PCAC relation
\beq
(m_u+m_d)\la \bar uu+\bar dd\ra = -2m_\pi^2 f_\pi^2 \,,\quad f_\pi = 93\,{\rm MeV},
\eeq
and we shall assume $\la \bar uu\ra\simeq \la\bar dd\ra {\equiv \la\bar qq\ra}$. 
\section{The lowest  Laplace sum rule moment at LO\label{sec:psi}}
Here, we study the dependence of the mass and coupling versus the external sum rule variable $\tau$ and QCD continuum threshold $t_c$.  

\begin{figure}[hbt]
\vspace*{-0.25cm}
\begin{center}
\hspace*{-5.5cm} { \boldmath\large${\cal R}^c_{10}$}\\
\includegraphics[width=9cm]{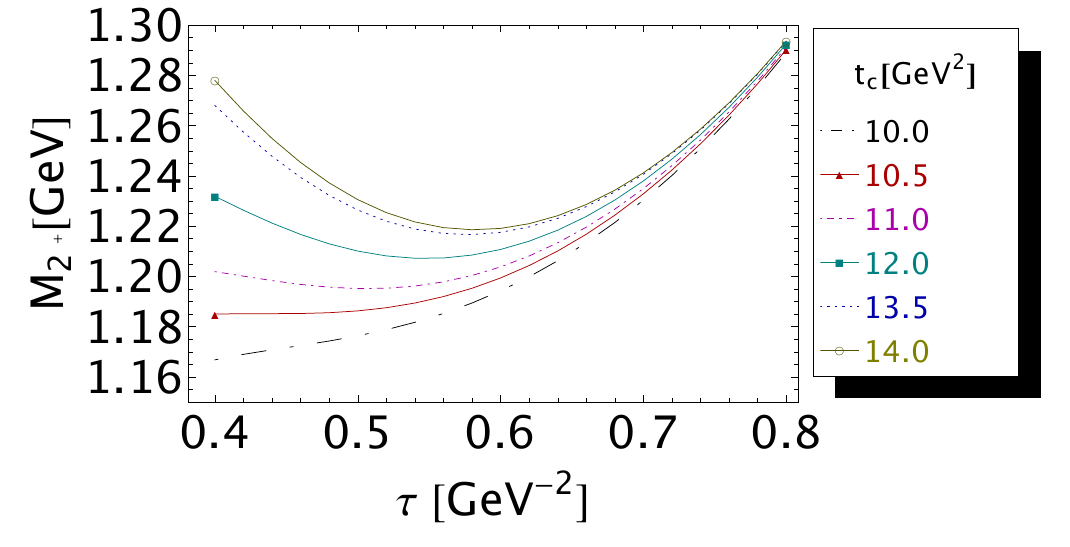}
\vspace*{-0.5cm}
\caption{\footnotesize  $\tau$-behaviour of the hybrid mass for different values of $t_c$ from ${\cal R}^c_{10}$. } 
\label{fig:mass}
\end{center}
\vspace*{-0.75cm}
\end{figure}

\begin{figure}[hbt]
\begin{center}
\hspace*{-5.5cm} { \boldmath\large${\cal L}^c_{0}$}\\
\includegraphics[width=9cm]{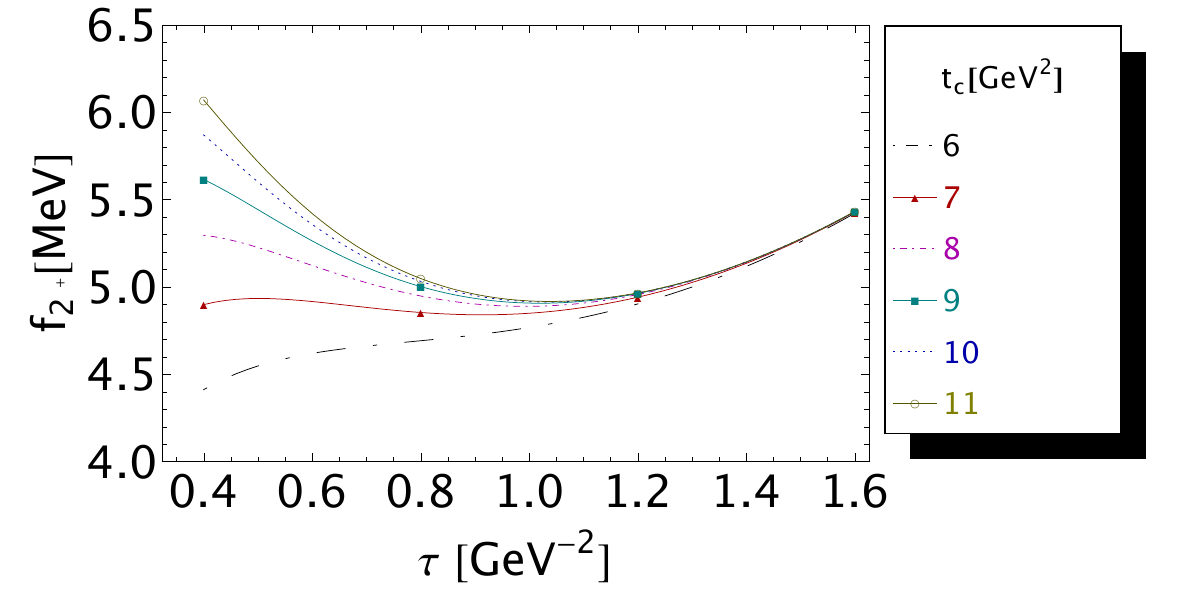}
\vspace*{-0.5cm}
\caption{\footnotesize  $\tau$-behaviour of the hybrid coupling for different values of $t_c$ from ${\cal L}_{0}^c$. } 
\label{fig:coupling}
\end{center}
\vspace*{-0.5cm}
\end{figure}

\d We consider as optimal the $t_c$ value obtained at the $\tau$-stability point (flat plateau, extremum, or inflexion point). The analysis is shown in Figs.\,\ref{fig:mass} and \ref{fig:coupling}. 

\d To be conservative, we consider the optimal result  in the range of $t_c$-values from the beginning of $\tau$-stability to the one where $t_c$-stability is obtained (lowest ground state dominance).

\d For the mass (resp.\ coupling), $\tau$-stability is obtained at $0.6$ GeV$^{-2}$ (resp.\ $1$ GeV$^{-2}$).   We obtain the results:
\bea 
\hspace*{-1cm}M_{2^+} &=& (1195\sim 1219)\,{\rm MeV}\,\, \,{\circ}{\hspace*{-0.3cm}\lrar} \, t_c=(11\sim 14)~{\rm GeV^2},\nnb\\
f_{2^+}&=&(4.85\sim 4.93) \, {\rm MeV} \, \,{\circ}{\hspace*{-0.3cm}\lrar} \,   t_c=(7\sim 11)~{\rm GeV^2}.
\label{eq:res0}
\eea
One can remark that the position of the optimal values for the mass and coupling are obtained at different $\tau$-values. This is due to different re-organization of the condensate contributions in the OPE for the moment and ratio of moments. One can also remark that the value of $t_c$ where the optimal values of the mass and coupling are simultaneously obtained is around 11 GeV$^2$ which one may eventually consider as the central value of the optimal results. The deviation from this central value can be considered as the systematics of the approach. 
\section{Higher Laplace sum rule moments at LO}
We may also consider higher moments :
\beq
{\cal L}_{n}^c(\tau,\nu)
=\int_{t>}^{t_c}dt~t^n e^{-t\tau}\frac{1}{\pi} \mbox{Im}\,\psi_{qg}(t,\nu)~~~{\rm for}~n=1,2,3
\eeq
and their ratios:
\beq
 {\cal R}^c_{n,n-1}(\tau)\equiv\frac{{\cal L}^c_{n}} {{\cal L}^c_{n-1}} ~~~{\rm for}~n=2,3.
  \label{eq:mom}
\eeq

\d {\it The highest moments ${\cal L}^c_{2,3}$ }

\begin{figure}[hbt]
\vspace*{-0.25cm}
\begin{center}
\hspace*{-5.5cm} { \boldmath\large${\cal R}^c_{32}$}\\
\includegraphics[width=9cm]{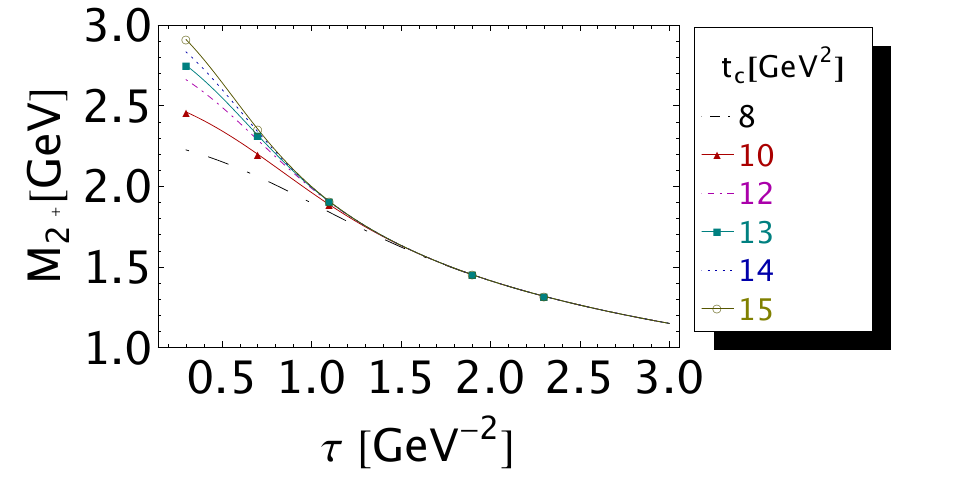}
\vspace*{-0.5cm}
\caption{\footnotesize  $\tau$-behaviour of the hybrid mass for different values of $t_c$ from the ratio of moments ${\cal R}^c_{32}$.
} 
\label{fig:mass1}
\end{center}
\vspace*{-0.5cm}
\end{figure}

\begin{figure}[hbt]
\vspace*{-0.5cm}
\begin{center}
\hspace*{-5.5cm} { \boldmath\large${\cal L}^c_{2}$}\\
\includegraphics[width=9cm]{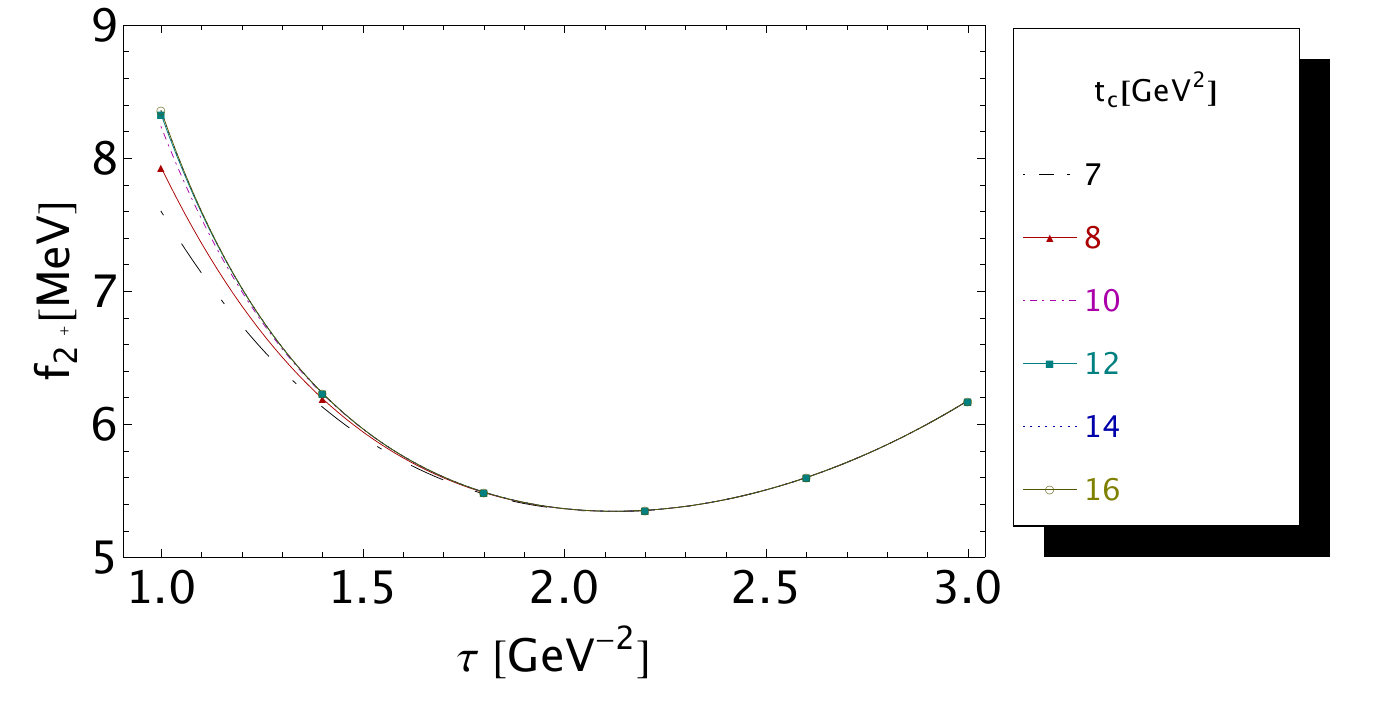}
\vspace*{-0.5cm}
\caption{\footnotesize  $\tau$-behaviour of the hybrid coupling for different values of $t_c$ from ${\cal L}^c_{2}$. 
} 
\label{fig:coupling1}
\end{center}
\vspace*{-0.5cm}
\end{figure}

These highest moments would correspond to the two-point function\,: 
\beq
 \Pi_{qg}(Q^2) = (Q^2)^2\psi_{qg}(Q^2),
 \eeq
obtained by applying, to the two-point correlator, the dimensionless projector 
\beq
\eta^{\mu\nu} \equiv g^{\mu\nu}-q^\mu q^\nu/q^2 = \tilde \eta^{\mu\nu} /q^2 .
\eeq
In this case, one can remark that, at LO only the log-terms in the OPE remain, after taking the Laplace transform
obtained  from the 2nd derivative in $\tau$ of ${\cal L}_{0}^c(\tau)$\,\footnote{Alternatively, one can derive ${\cal L}_{2}^c(\tau)$ by taking the Laplace transform of the 4th derivative in $Q^2$ of $ \Pi_{qg}(Q^2)$
which is superconvergent and obeys an homogeneus RGE.}:
\beq
{\cal L}_{2}^c(\tau)\vert_{\rm LO}=  \tau^{-4}\Big{\{} 6d_0(1-\rho_4)-2d_2\tau+d_{4q}\tau^2-d_{6qg}\tau^3\Big{\}}.
\label{eq:Laplace_Pi}
\eeq
However, these contributions vanish in the chiral limit.  
The analysis is shown in Figs.\,\ref{fig:mass1} and \ref{fig:coupling1}. 

One can notice that unlike ${\cal R}^c_{10}$, the analysis of the mass from   ${\cal R}^c_{32}$
shows neither a $\tau$-minimum nor a clear inflexion point. Using the value of $M_{2^+}$ found in Eq.\,\ref{eq:res0}, $\tau$-minimas of \ta{${\cal L}_{2}^c$} 
are found around $\tau\simeq 2 $ GeV$^{-2}$, at which we extract\,:
\beq
f_{2^+}=(5.23\sim 5.35) \,{\rm MeV} \, \,{\circ}{\hspace*{-0.3cm}\lrar} \,   t_c=(4.5\sim 6)~{\rm GeV^2}. 
\label{eq:res-pi}
\eeq
which is consistent with the one in Eq.\,\ref{eq:res0} but could be less reliable because it is obtained at higher $\tau$-values. 

\d {\it Ratios  of intermediate and  higher moments }

Using the MDA, we use different ratios of higher moments
\beq
{\cal R}^c_{21},{\cal R}^c_{31},\,{\cal R}^c_{32},\,
\eeq
to re-extract the ground state mass. The moment ${\cal L}_1$ (resp. ${\cal L}_3$) is obtained from the $\tau$-derivative of ${\cal L}_0$ (resp. ${\cal L}_2$).
The LO analysis shows curves having the same shape as in Fig.\,\ref {fig:mass1} 
such that the results are not conclusive. 

For a further use, we give below the QCD expression of  ${\cal L}_1^c(\tau)$\,: 
\bea
{\cal L}_{1}^c(\tau)\vert_{\rm LO}&=&  \tau^{-3}\Big{\{} 2d_0(1-\rho_3)-d_2\tau+d_{4q}\tau^2\nnb\\
&&+\big{[}\gamma_Ed_{6qg}-(d_{6q}+d_{6g}+d^c_{6qg})\big{]}\tau^3\Big{\}},
\label{eq:L2-lo}
\eea

\section{The two-point function at NLO in PT\label{sec:nloPT}}
\begin{figure}[!h]
\begin{center}
\includegraphics[width=7.75cm]{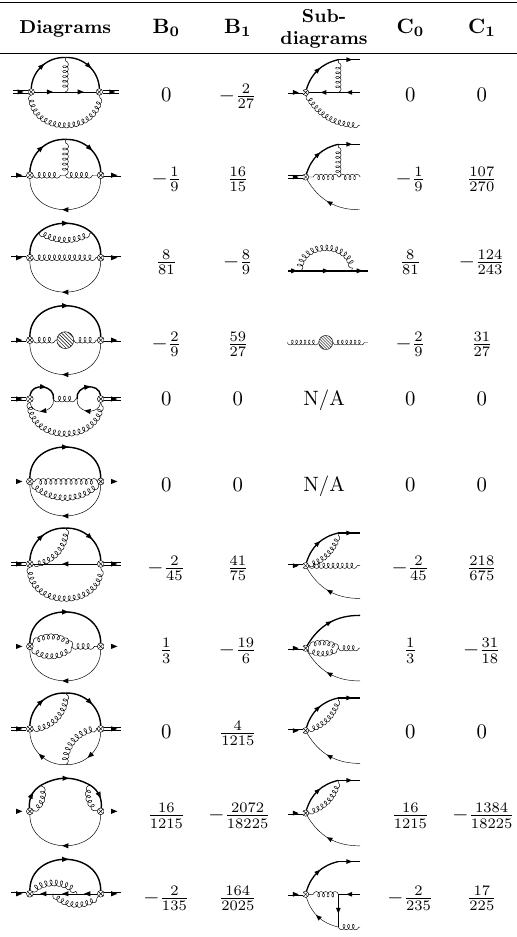}
\vspace*{-0.3cm}
\caption{\footnotesize  Contributions to the bare two-point correlator at NLO of PT series. From top to bottom, we label the diagrams as $\psi_B^{(i)}$, $i\in\left\{1,2,\ldots,11\right\}$. The numeric coefficients are defined in Eqs.~\ref{eqn:nlo_pi_b} and \ref{eqn:nlo_pi_c}. Topologically equivalent permutations are not shown. The shaded loop represents the gluon vacuum polarization, i.e., the sum of a quark loop, a gluon loop, and a ghost loop. All other notations are identical to Fig.~\ref{fig:ope}. Feynman diagrams were drawn using TikZ-Feynman~\cite{Ellis:2016jkw}. } 
\label{fig:opeNLO}
\end{center}
\vspace*{-0.5cm}
\end{figure}

NLO Feynman gauge PT contributions are calculated in the chiral limit using the $\overline{\rm MS}$ scheme with $n=4+2\epsilon$ and diagrammatic renormalization methods~\cite{deOliveira:2022eeq}. NLO Feynman diagrams are shown in Fig.~\ref{fig:opeNLO}. 

FeynCalc~\cite{Mertig:1990an,Shtabovenko:2016sxi,Shtabovenko:2020gxv,Shtabovenko:2023idz}, Tarcer~\cite{Mertig:1998vk}, and LiteRed~\cite{Lee:2012cn,Lee:2013mka} are used. All loop integrals can be evaluated using results given in Ref.~\cite{Pascual_and_Tarrach}. The bare NLO PT correlator contributions are parametrized as 
\beq
\psi_B^{(i)}\left(Q^2\right) = \frac{Q^2 a_s^2}{256\pi^2} \left[ B_0^i \frac{L}{\epsilon} + B_1^i L + \frac{3}{2} B_0^i L^2 \right] \,. 
\label{eqn:nlo_pi_b}
\eeq
Following the methodology of Ref.~\cite{deOliveira:2022eeq}, the divergent parts sub-diagrams shown in Fig.~\ref{fig:opeNLO} are used to construct counter-terms, parametrized as 
\beq
\psi_C^{(i)}\left(Q^2\right) = \frac{Q^2 a_s^2}{256\pi^2} \left[ C_0^i \frac{L}{\epsilon} + C_1^i L + C_0^i L^2 \right] \,. 
\label{eqn:nlo_pi_c}
\eeq
Note that $n_f=3$ is used in Fig.~\ref{fig:opeNLO}; however, for general $n_f$ and $i=4$, $B_0=2\left(2n_f-15\right)/81,$ $B_1=\left(861-110n_f\right)/243$, $C_0~=~B_0$, and $C_1=-31\left(2n_f-15\right)/243$. For diagrams 
$\psi_{\rm B}^{(i)}$, 
$i \in \{1, 2, 9, 10, 11\}$, $C_0$ and $C_1$ in Fig.~\ref{fig:opeNLO} include a factor of 2 to account for independent (but equal) sub-diagrams and their associated counter-term diagrams.

The renormalized NLO PT contribution is
\beq
\begin{split}
&\psi^{\rm pert}_{qg}\!\left.\right|_{NLO}\left(Q^2\right) = \sum_{i=1}^{11} M_{(i)}\left[\psi_B^{(i)}-\psi_C^{(i)}\right]
= Q^2\left[ d_0^\prime L + \tilde{d}_0 L^2 \right] \,,
\\
&d_0^\prime = \frac{a_s^2}{864\pi^2}\left(\frac{317-720n_f}{1080}\right)
\overset{n_f=3}{=}\frac{ a_s^2}{864\pi^2}\left(-\frac{1843}{1080}\right)  \,,
\\
&\tilde{d}_0 =  \frac{a_s^2}{864\pi^2}\left(\frac{11+6n_f}{72}\right)\overset{n_f=3}{=}
\frac{a_s^2}{864\pi^2}\left(\frac{29}{72}\right)  \,,
\end{split}
\label{eqn:nlo_pi_renorm}
\eeq
where the multiplicity of each diagrams is $M_{(i)}=1$ for $i \in \{1, 4,5,6 \}$, $M_{(i)}=2$ for $i \in \{2,3,8,9,10,11\}$, and $M_{(7)}=4$.

The complete dimension-zero NLO contribution to the PT series is
\beq
\begin{split}
&\psi_{qg}^{\rm pert}\left(Q^2\right)= Q^2 \left[ a_s a_{00}L+a_s^2\left(a_{10}L+a_{11} L^2\right)   \right] \,, 
\\
&a_{00}=\frac{d_0}{a_s}\,, \quad a_{10}= \frac{d_0^\prime}{a_s^2} \,, \quad a_{11}= \frac{\tilde{d}_0}{a_s^2}\,, 
\end{split}
\label{eqn:a_terms}
\eeq
where $d_0$ defined in Eq.~\ref{eq:ope} can be deduced from Fig\,\ref{fig:ope} while $d_0^\prime$ and $\tilde{d}_0$ are given in Eq.~\ref{eqn:nlo_pi_renorm}. Note that for $n_f=3$, $a_{10}/a_{00}=-1.706$ and  $a_{11}/a_{00}=0.403$ indicating that radiative contributions are meaningful while still being well-controlled. 
\section{NLO leading-log contributions of the condensates\label{sec:nloNP}}
{As evident from Eq.\,\ref{eq:Laplace_Pi}, the higher weight ($n\ge 2$) 
Laplace sum-rules} only have NP dependence on $d_{4q}$ and $d_{6qg}$, so their NP content is limited. Thus it is desirable to determine NLO leading logarithmic corrections to the $\langle \alpha_s G^2 \rangle$, $\langle g_s^3 G^3\rangle$, and $\alpha_s \langle \bar{u}u \rangle \langle \bar{d} d \rangle$  
condensate contributions 
because, similar to the LO mixed and quark condensates, these logarithmic corrections will contribute to 
{the higher weight ($n\ge 2$) 
Laplace sum-rules in Eq.\,\ref{eq:Laplace_Pi}}.  Such leading-logarithm condensate contributions can be determined via renormalization group equation (RGE) methods, a procedure that has previously been applied to tensor gluonium \cite{SLI}.  This procedure requires the correlator anomalous dimension which can be obtained from our NLO PT results. In the chiral limit, the 
contributions to the Adler-type correlation function satisfy a homogeneous RGE of the form\,: 
\begin{equation}
\left[\nu\frac{\partial}{\partial \nu}+\beta\, a_s\frac{\partial }{\partial a_s} -2\gamma^\psi\right]
\frac{d^2\psi^{\rm pert}}{d\left(Q^2\right)^2} =0  
\label{eqn:general_RG}
\end{equation}
where $\gamma^\psi$ is the anomalous dimension for the correlation functions $\psi_{qg}$ 
emerging from renormalization of the composite operator Eq.\,\ref{eq:current}. Using the LO expansions of the $\beta$ function and anomalous dimension\,:
\begin{equation}
  \beta=\beta_1 \frac{\alpha_s}{\pi}+\ldots\, ,
  ~\gamma^\psi=\gamma_1^\psi \frac{\alpha_s}{\pi}+\ldots  
\label{eq:RG_beta_gammapsi}
\end{equation} 
and following the methodology of Ref.~\cite{SLI}, we determine the LO anomalous dimension $\gamma_1^\psi$ to be\,:
\begin{equation}
   \gamma_1^\psi=
   \frac{\beta_1}{2}-2\frac{a_{11}}{a_{00}} =
   -\frac{55}{18}
   \,.
   \label{eq:gamma1psi}
\end{equation} 

A general RGE analysis can be formulated for the $\langle \alpha_s G^2 \rangle$, $\langle g_s^3 G^3\rangle$, and $\alpha_s \langle \bar{u}u \rangle \langle \bar{d} d \rangle$ terms, which we parametrize as\,: 
\beq
\begin{split}
&\psi^{\alpha_s G^2}\left(Q^2\right)=
\frac{1}{Q^2}
\langle \alpha_s G^2 \rangle\left[
b_{00}+ a_s\left( b_{10}+b_{11}L \right)\right] \,,
\\
&\psi^{g^3 G^3}\left(Q^2\right)=\frac{1}{Q^4}\langle g_s^3 G^3\rangle 
\left[
c_{00}+ a_s\left( c_{10}+c_{11}L \right)\right]\,,
\\
&\psi^{\alpha \bar{q}q \bar{q}q}\left(Q^2\right)=\frac{1}{Q^4}
\rho\alpha_s \langle \bar{u}u \rangle \langle \bar{d} d \rangle
\left[
e_{00}+ a_s\left( e_{10}+e_{11}L \right)\right] \,,
\\
&b_{00}=\frac{1}{72\pi}\,,~c_{00}=\frac{1}{216\pi^2}\,,~e_{00}=-\frac{2\pi}{9}\,.
\label{eqn:b_c_d_terms}
\end{split}
\eeq
Because these LO contributions are finite, the NLO condensate contributions Eq.\,\ref{eqn:b_c_d_terms} satisfy a homogeneous RGE identical in form to Eq.\,\ref{eqn:general_RG} with the Adler function replaced by the condensate expressions in Eq.\,\ref{eqn:b_c_d_terms}. For a generic condensate contribution of the form\,: 
\begin{equation}
 \psi^\chi
 =\langle \chi \rangle \left[ 
 E+a_s\left( F+H\, L \right)
 \right]  \, 
\end{equation}
where at LO there is no mixing in the anomalous dimension for the operator $\chi$\,:
\begin{equation}
    \nu\frac{d}{d\nu}\langle \chi \rangle=-\gamma_1^\chi \, a_s\, \langle \chi \rangle\,,
\end{equation}
the RGE methodology of Ref.~\cite{SLI} gives the solution\,:
\begin{equation}
H=-\frac{1}{2}\left(2\gamma_1^\psi+\gamma_1^\chi  \right)E \,.
\label{eq:chi_RG_soln}
\end{equation}
A summary of the condensates and corresponding leading-log coefficients $H$ is shown in Table~\ref{tab:nll-condensates}.
\begin{table}
\setlength{\tabcolsep}{0.3pc}
    \centering
    \begin{tabular}{ccc}
    \hline
       $\langle \chi \rangle$ & $\gamma_1^\chi$ & $H$
       \\[3pt] 
       \hline
        $\langle \alpha_s G^2 \rangle$ & $\gamma_1^{\alpha_s G^2}=0$ & $b_{11}=-\gamma_1^\psi b_{00}$\\[3pt]
        $\rho\alpha_s \langle \bar{u}u \rangle \langle \bar{d} d \rangle$ & $\gamma_1^{\alpha\bar q q \bar q q}  =-\beta_1-2\gamma_1$ & $e_{11}=\left(-\gamma_1^\psi+\frac{\beta_1}{2}+\gamma_1^m \right)e_{00}$
        \\[3pt]
        $\langle g_s^3 G^3\rangle$ & $\gamma_1^{g^3G^3}  =\frac{7}{2}+\frac{n_f}{3}-\beta_1$ & $ c_{11}=  -\left(\gamma_1^\psi+\frac{1}{2}\gamma_1^{g^3G^3}\right) c_{00}$
    \end{tabular}
    \caption{Leading-Log condensate corrections for general $n_f$. The vacuum saturation approximation is implicit in $e_{11}$. The anomalous dimension $\gamma_1^{g^3G^3}$ was determined by generalizing $n_f=1$ results from Ref.~\cite{Narison:1983kn}} 
    \label{tab:nll-condensates}
\end{table}

The final $n_f=3$ expressions for the leading-logarithm NLO corrections to the  $\langle \alpha_s G^2 \rangle$, $\langle g_s^3 G^3\rangle$, and $\alpha_s \langle \bar{u}u \rangle \langle \bar{d} d \rangle$ terms are\,:
\begin{equation}
\begin{split}
&b_{11}=
{\frac{55}{1296\pi}}
\,,\quad c_{11}=
{-\frac{13}{1944\pi^2}} \,,
\\
&e_{11}={-\frac{(83+6n_f)\pi}{162}\overset{n_f=3}{=}-\frac{101\pi}{162} }
\,.
\end{split}
\label{eq:b11_c11_e11}
\end{equation}
We note that the RGE methodology only determines the leading-logarithmic terms, and hence the quantities $b_{10}$, $c_{10}$, and $e_{10}$ cannot be determined by RGE methods. 
Including the leading-logarithmic contributions of the condensate in Eq.\,\ref{eq:b11_c11_e11}, the total NLO corrections to the $\psi_{qg}$ correlation function at order $a_s^2$ and in the chiral limit are\,:
\beq
\psi_{qg}\vert_{\rm NLO}\left(Q^2\right) =Q^2 \left[ \tilde{d}_0 L^2 + d_0^\prime L
+\left(\frac{\tilde{d}_{4g}}{Q^4}
+\frac{\tilde{d}_{6g}+\tilde{d}_{6q}}{Q^6}\right)L\right]
\eeq
\bea
&\tilde{d}_{4g} = a_s\frac{55}{1296\pi}\langle \alpha_s G^2 \rangle \,,
\quad
\tilde{d}_{6g} =-a_s \frac{13}{1944\pi^2} \langle g^3G^3\rangle\nnb \,,
\\
&\tilde{d}_{6q} =-a_s \frac{(83+6n_f)\pi}{162}
\rho\alpha_s \langle \bar{u}u \rangle \langle \bar{d}d \rangle,  
\label{tilded6q}
\eea
where $d_0^\prime$, $\tilde{d}_0$ are given in Eq.~\ref{eqn:nlo_pi_renorm}. 

\section{Laplace sum rules analysis at NLO }
To the LO expression in Eq.\,\ref{eq:srlo}, we add the leading-log. NLO corrections obtained in previous sections\,\ref{sec:nloPT} and \ref{sec:nloNP} in the chiral limit $m_q=0$\,:
\bea
\hspace*{-0.5cm}
{\cal L}_{0}^c(\tau)\vert_{\rm NLO}&=&\tau^{-2}\Bigg{\{} \Big{[}2\tilde d_0(1-\gamma_E) +d'_0\Big{]}(1-\rho_2)-\tilde d_{4g}\gamma_E\tau^2\nnb\\
&&+\,(\tilde d_{6g}+\tilde d_{6q})(1-\gamma_E)\tau^3\Bigg{\}},
\label{eq:srnlo}
\eea
{\d \it Mass and coupling of the ground state from ${\cal R}_{10}^c$ and ${\cal L}_{0}^c$}

\begin{figure}[hbt]
\vspace*{-0.25cm}
\begin{center}
\hspace*{-5.5cm} { \boldmath ${\cal R}_{10}^c$}\\
\includegraphics[width=8.5cm]{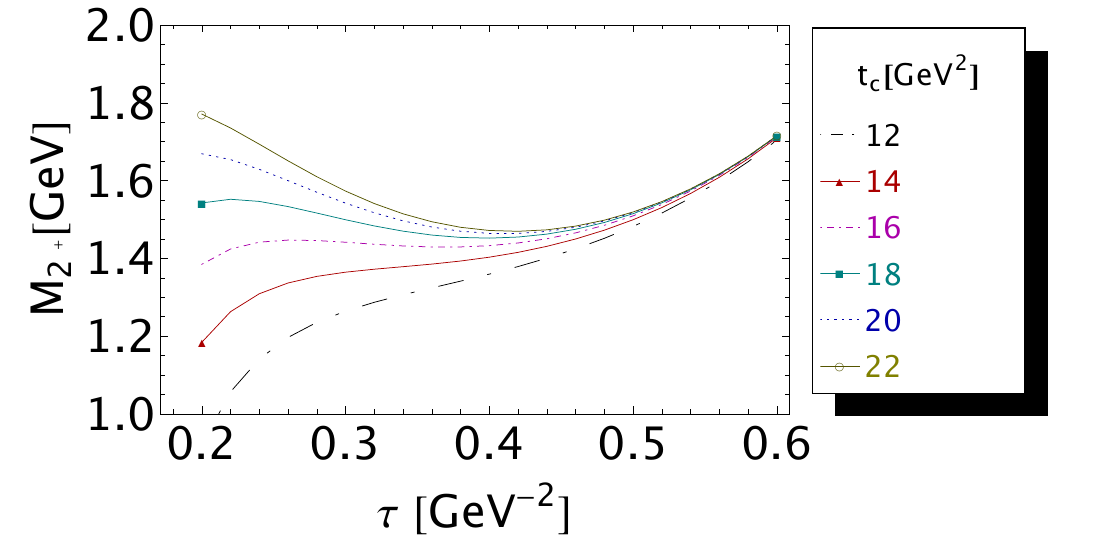}\\
\hspace*{-5.5cm} { \boldmath ${\cal L}_{0}^c$}\\
\includegraphics[width=8.5cm]{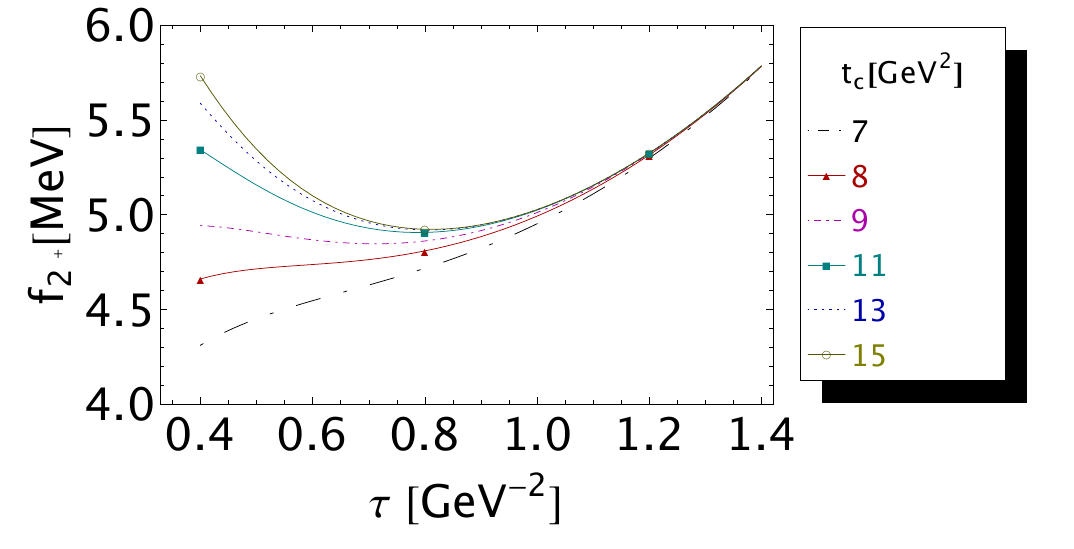}
\vspace*{-0.5cm}
\caption{\footnotesize $\tau$-behaviour of the lowest ground state parameters at NLO\,: mass from ${\cal R}_{10}^c$ and coupling from ${\cal L}_{0}^c$ for different values of $t_c$. } 
\label{fig:2+nlo}
\end{center}
\vspace*{-0.5cm}
\end{figure}

We repeat the analysis in section\,\ref{sec:psi} by adding the NLO corrections. The results
are shown in Fig.\,\ref{fig:2+nlo}.
For the mass (resp.\ coupling), $\tau$-stability is obtained at $0.4$ GeV$^{-2}$ (resp.\ $0.8$ GeV$^{-2}$), at which we extract
\bea 
\hspace*{-1cm}M_{2^+} &=& (1393\sim 1469)\,{\rm MeV}\,\, \,{\circ}{\hspace*{-0.3cm}\lrar} \, t_c=(14\sim 22)~{\rm GeV^2},\nnb\\
f_{2^+}&=&(4.85\sim 4.92)~ {\rm MeV} \, \,{\circ}{\hspace*{-0.3cm}\lrar} \,   t_c=(9\sim 16)~{\rm GeV^2}.
\label{eq:res0-nlo}
\eea
{\d \it Mass of coupling of the ground state from {${\cal R}_{32}^c$ and ${\cal L}_{2}^c$}}

The LO expression of {${\cal L}_2^c$} is given in Eq.\,\ref{eq:Laplace_Pi} to which we add the NLO expression obtained in the chiral limit\,:

\bea
{\cal L}_{2}^c(\tau)\vert_{\rm NLO}&=&  \tau^{-4}\Bigg{\{} \Bigg{[} \ga 6d'_0+22(1-\frac{6}{11}\gamma_E)\tilde d_0\dr\Bigg{]}(1-\rho_4)\nnb\\
&&+\tilde d_{4g}\tau^2
-(\tilde d_{6q}+\tilde d_{6g})\big{]}\tau^3\Bigg{\}}. 
\label{eq:L2-nlo}
\eea

We attempt to re-determine the mass from the highest unsubtracted ratio of moments ${\cal R}_{32}^c$.
The shape of the curves is similar to the one  in Fig.\,\ref{fig:mass1} which does not show a clear
inflexion point. Unfortunately,  the addition of the NLO corrections does not help for stabilizing this ratio of high moments.

\begin{figure}[H]
\vspace*{-0.25cm}
\begin{center}
\hspace*{-5.5cm} { \boldmath ${\cal L}_{2}^c$}\\
\includegraphics[width=8.5cm]{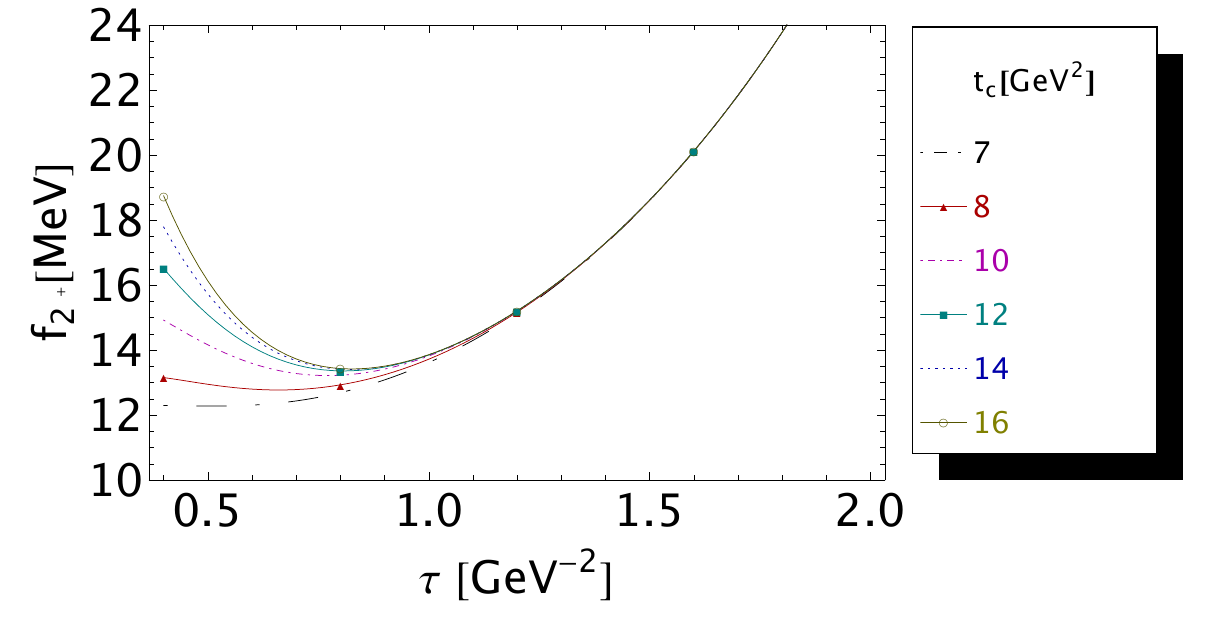}
\vspace*{-0.5cm}
\caption{\footnotesize : $\tau$-behaviour of the lowest ground state coupling at NLO for different values of $t_c$ from ${\cal L}_{2}^c$. } 
\label{fig:L2-nlo}
\end{center}
\vspace*{-0.5cm}
\end{figure}

We repeat the analysis for extracting $f_{2^+}$ from ${\cal L}_{2}^c$ by including NLO corrections and using as input the meson mass determined at NLO in Eq.\,\ref{eq:res0-nlo}. We show the analysis in Fig.\,\ref{fig:L2-nlo} and deduce the optimal result at the minimum $\tau\simeq 0.8$ GeV$^{-2}$\,:
\beq
f_{2^+}=(12.86\sim 13.51)~ {\rm MeV} \, \,{\circ}{\hspace*{-0.3cm}\lrar} \,   t_c=(8\sim 16)~{\rm GeV^2}.
\label{eq:res2-nlo}
\eeq
This result is about 2.4 times the LO one in Eq.\,\ref{eq:res-pi} indicating that the effect of NLO corrections are huge for this moment. This is mainly due to the new contributions of the gluon and four-quark condensates absent at LO. This result can be improved by the inclusion of higher order $\alpha_s$ corrections in the non-perturbative contributions.  
For a conservative result, we consider the mean from ${\cal L}_{0}^c$ in Eq.\,\ref{eq:res0-nlo} and  ${\cal L}_{2}^c$ in Eq.\,\ref{eq:res2-nlo}  and take as a systematics the deviation of the two determinations from the central value. We obtain\,:
\beq
\la f_{2^+}\ra  =9.04(0.65)(4.15)_{syst}=(9.04\pm 4.2)~ {\rm MeV}.
\label{eq:f2-nlo}
\eeq
The first error is the mean of the sum of the errors from the two different determinations of $f_{2^+}$ given in Table\,\ref{tab:error} while the second is the   systematics. 
\section{The quest for the $2^{++}$  hybrid topological charge $\Pi_{qg}(0)$}

We have ignored the eventual effect of the two-point function subtraction constant at zero momentum (hybrid topological charge) $\Pi_{qg}(0)$  in the analysis of the $\psi$-sum rule.  However, if $\psi_{qg}(q^2\equiv -Q^2)$ can be considered 
as $\Pi_{qg}(q^2)/q^4$,  the corresponding subtracted dispersion relation used for  $\psi_{qg}(Q^2)$ should be affected by $\Pi_{qg}(0)$ and its slope $\Pi'_{qg}(0)$\,\cite{SNV,SNG}:
\bea
\psi(q^2)&\equiv& \frac{\Pi_{qg}(q^2)-\Pi_{qg}(0)}{q^4}  +\frac{\Pi'_{qg}(0)}{q^2} +\cdots\nnb\\
&=&\int_0^\infty \frac{dt}{t^2(t-q^2-i\epsilon)}\frac{1}{\pi}\,{\rm Im} \Pi_{qg}(t)
\eea

 
 We check the previous result by  the estimate of $\Pi_{qg}(0)$  using the ${\cal L}^c_1$ moment following Refs.\,\cite{SNG,SNV} for the scalar gluonium and Refs.\,\cite{SNG0,SNG,SHORE} for the topological charge.  Neglecting (to a first approximation) $\Pi'_{qg}(0)$, the sum rule reads: 
 \beq
 \Pi_{qg}(0)=  2f_2^2M_2^4\,e^{-M_2^2\tau} - {\cal L}^c_1(\tau)\vert_{\rm QCD}. 
\eeq
The LO expression of ${\cal L}_1^c$ is given in Eq.\,\ref{eq:L2-lo} to which we add the NLO expression obtained in the chiral limit\,:
\bea
{\cal L}_{1}^c(\tau)\vert_{\rm NLO}&=&  \tau^{-3}\Big{\{} \big{[} 2d'_0+6\tilde d_0(1-\frac{2}{3}\gamma_E)\big{]}(1-\rho_3)\nnb\\
&&+\tilde d_{4g}\tau^2
+\big{[}\gamma_E(\tilde d_{6q}+\tilde d_{6g})\big{]}\tau^3\Big{\}}.
\label{eq:l2-nlo}
\eea

\begin{figure}[H]
\begin{center}
\includegraphics[width=8.5cm]{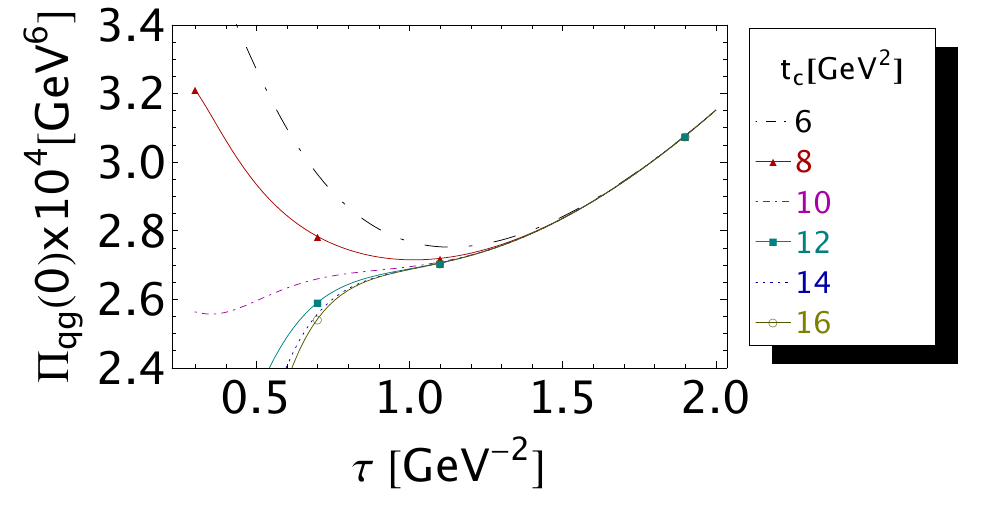}
\vspace*{-0.5cm}
\caption{\footnotesize $\tau$-behaviour of the subtraction constant $ \Pi_{qg}(0)$ at NLO for different values of $t_c$. } 
\label{fig:psi0}
\end{center}
\vspace*{-0.5cm}
\end{figure} 

The analysis is shown in Fig.\,\ref{fig:psi0} where the $\tau$-stability is obtained for $\tau\simeq (1.0\sim 1.1)$ GeV$^{-2}$.  We deduce for $t_c=(6\sim 16)~{\rm GeV}^2$\,:

\beq
  \Pi_{qg}(0)\simeq  2.72\ \times 10^{-4}\,{\rm GeV}^6\,.
 \label{eq:pi0}
 \eeq

 \vspace*{-0.5cm}
 {\section{Improved estimate of  $\Pi_{qg}(0),\, M_{2^+}$ and $f_{2^+}$}}
\vspace*{-0.25cm}
\begin{table}[H]
\setlength{\tabcolsep}{0.22pc}
\begin{center}
    {\small
\begin{tabular}{cccccccccc}
\hline
&$\tau$&$t_c$&$\alpha_s$&$\la\alpha_s G^2\ra$&$\la\bar\psi G\psi\ra$&$\la G^3\ra$&$\la \bar\psi\psi\ra^2$&$M_2$&$f_2$\\
\hline
$\Delta M_2$&30&47&125&95&0&0&92&--&--\\
$\Delta f_2^{{\cal L}_0}$&0.44&0.40&0.04&0.00&0.00&0.00&0&0.40&--\\

$\Delta f_2^{{\cal L}_2}$&0&0.16&0.17&0.19&0&0.08&0.11&1.11&--\\
$\Delta\Pi_{qg}(0)$&0&0.02&0.00&0.18&0.00&0.26&0.26&0.07&0.12\\
\hline
\end{tabular}
}
\end{center}
\vspace*{-0.5cm}
    \caption{
Different sources of errors.  $\Delta f_2$ and $\Delta M_2$ are in units of MeV.  $\Delta\Pi_{qg}(0)$ is in units of $ 10^{-4}$   GeV$^6$.} 
    \label{tab:error}
    \vspace*{-0.25cm}
\end{table}
We improve the previous estimates by including $\Pi_{qg}(0)$ into the expression of ${\cal L}_0^c$ and get the modified sum rule\,:
\beq
{\cal L}_0^c +\tau\,\Pi_{qg}(0)=2f_2^2M_2^2\,e^{-M_2^2\tau}.
\eeq

\begin{figure}[H]
\begin{center}
\includegraphics[width=8.5cm]{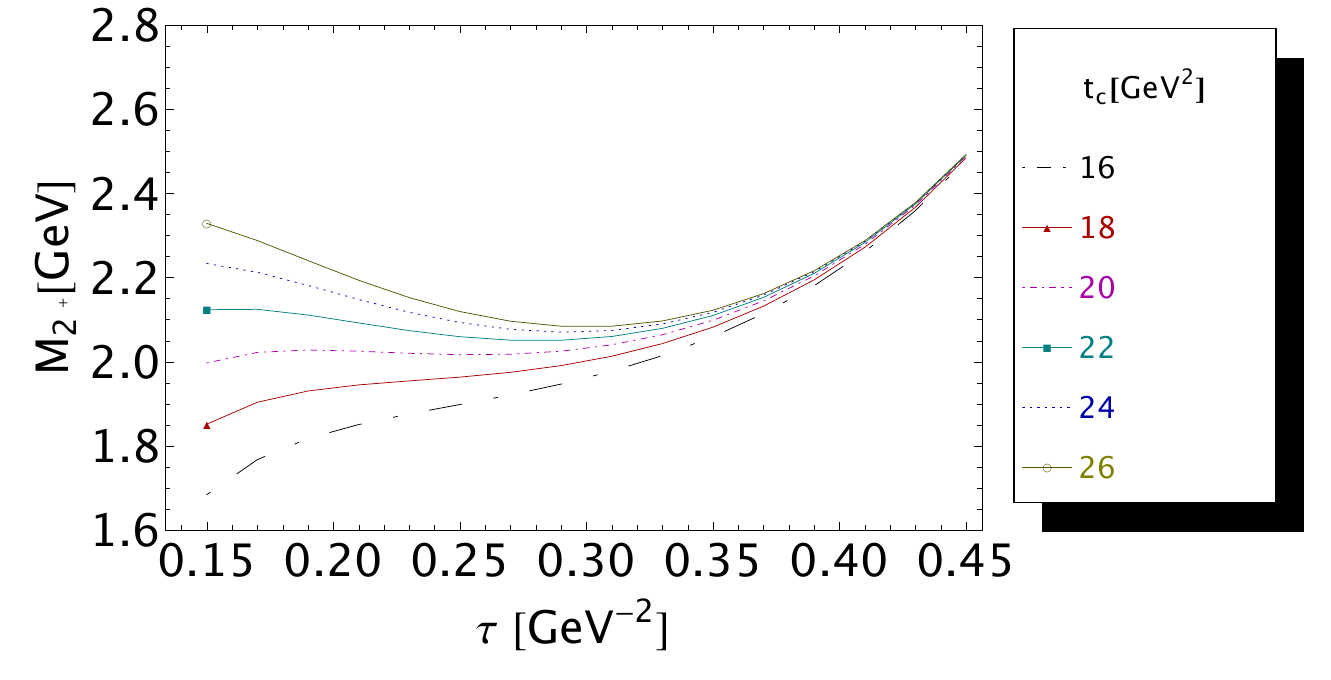}
\vspace*{-0.5cm}
\caption{\footnotesize $\tau$-behaviour of the improved $2^{++} $ hybrid mass at NLO for different values of $t_c$ after 3 iterations from ${\cal R}_{10}^c$. } 
\label{fig:mass-L0}
\end{center}
\vspace*{-0.5cm}
\end{figure}

\begin{figure}[H]
\begin{center}
\includegraphics[width=8.5cm]{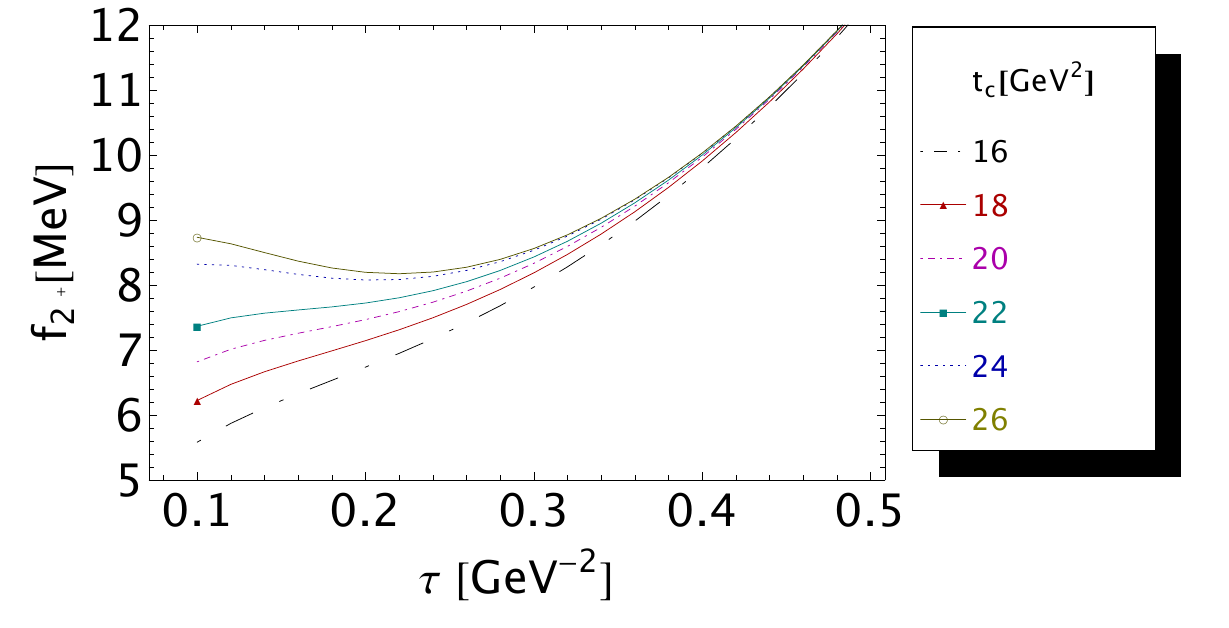}
\vspace*{-0.5cm}
\caption{\footnotesize $\tau$-behaviour of the improved $2^{++} $ hybrid coupling at NLO for different values of $t_c$ after 3 iterations from ${\cal L}_0^c$. } 
\label{fig:coupling-L0}
\end{center}
\vspace*{-0.5cm}
\end{figure}

\begin{figure}[H]
\begin{center}
\includegraphics[width=8.5cm]{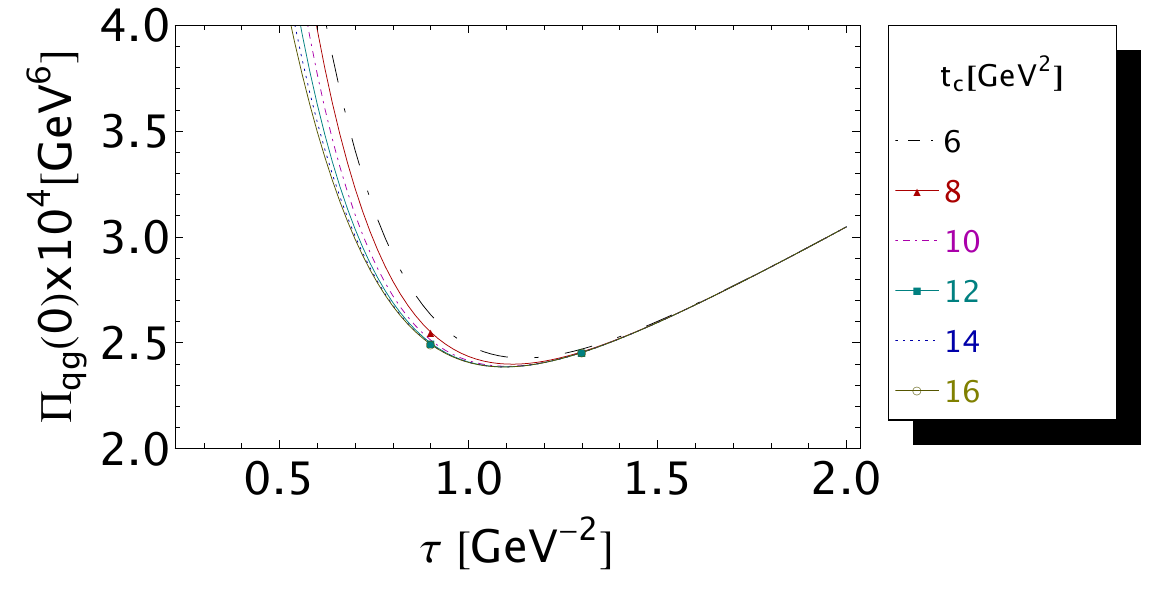}
\vspace*{-0.5cm}
\caption{\footnotesize $\tau$-behaviour of the improved subtraction constant $ \Pi_{qg}(0)$ at NLO for different values of $t_c$ after 3 iterations. } 
\label{fig:pi0}
\end{center}
\vspace*{-0.5cm}
\end{figure}

 Using as input, in the 1st iteration, the central value of  
$\Pi_{qg}(0)$ in Eq.\,\ref{eq:pi0} into ${\cal L}_0^c$ and ${\cal R}_{10}^c$, we extract the set $(f_{2^+},\, M_{2^+})$. Then,
we use the new mean value of $f_{2^+}$ from ${\cal L}_0^c$ and from ${\cal L}_2^c$ and the new value of $M_{2^+}$ to re-extract $\Pi_{qg}(0)$. We continue the procedure and find that the values of $(f_{2^+},\, M_{2^+})$ and 
$\Pi_{qg}(0)$ stabilize after 3 iterations. The analysis from the 3rd iteration is shown in Figs.\,\ref{fig:mass-L0},\, \ref{fig:coupling-L0} and \ref{fig:pi0}.  Including the different sources of errors given in Table\,\ref{tab:error}, we deduce the improved $2^{++}$ hybrid meson mass and coupling at NLO\,:
\bea
M_{2^+}&=&2038(190 )\,{\rm MeV},\nnb\\
\la f_{2^+}\ra&=&10.47(0.89)(2.73)_{syst}=10.47(2.87)\,{\rm MeV}, 
\label{eq:2plus}
\eea

where $\la f_{2^+}\ra $ is the mean value of $f_{2^+}$ from ${\cal L}_0^c$  and ${\cal L}_2^c$. The 2nd error is the departure from the mean central value. The improved value of the topological charge is\,:
\beq
\Pi_{qg}(0)=(2.41\pm 0.43)\times10^{-4}\, {\rm GeV}^6
\label{eq:pi02}
\eeq

 We notice that, in this $2^{++}$ channel,  the contribution $\Pi_{qg}(0)$ is important as it reduces the difference of value of $f_{2^+}$  from ${\cal L}_0^c$ and ${\cal L}_2^c$. This is not the case for the vector and scalar light hybrid channels where the effect of $\Pi_{qg}(0)$ is negligible\,\cite{BALITSKY}. 

\section{Comments and confrontation with the data}
\d Comparing the LO analysis of the lowest ground state mass and coupling shown in Figs.\,\ref{fig:mass} and \ref{fig:coupling}
with the one including NLO corrections and $\Pi_{qg}(0)$ shown in Figs.\,\ref{fig:mass-L0} and \ref{fig:coupling-L0}
one can notice that these contributions shift the position of the $\tau$-minimas from  $(0.6\sim 1.0)$\, to about 0.3 GeV$^{-2}$, where the OPE is more convergent . 

\d The  NLO corrections and especially the $\Pi_{qg}(0)$ contribution reduce the discrepancy between the estimate of the coupling  $f_{2^+}$  from the moment ${\cal L}_0^c$ and ${\cal L}_2^c$ at NLO. The important contribution of  $\Pi_{qg}(0)$ in the $2^{++}$ channel has to be contrasted with the one in the vector and scalar channels which are negligible according to Ref.\,\cite{BALITSKY}. 

\d The central value of the $2^{++}$ mass is about 400 MeV smaller than the one of lowest mass  prediction obtained in Ref.\,\cite{TAN} at LO from another choice of the interpolating current. However, we notice that the results of Ref.\,\cite{TAN} do not satisfy the $(\tau,t_c)$ stabilty criteria as shown in the example in Figs. 4 and 5 of their paper where the value of the mass increases with the continuum threshold $t_c$ and with the sum rule variable $1/\tau\equiv M^2$. This behaviour is similar to our ${\cal R}_{32}$ ratio of moment shown  in Fig.\,\ref{fig:mass1} of our paper from which we cannot extract any reliable prediction. One should note that our moment
${\cal L}_2^c$ has the same degree  as the one used in Ref.\,\cite{TAN} . Our analysis also differs from Ref.\,\cite{TAN} by the inclusion of NLO corrections in the PT and 
NLO leading-logarithms corrections due to the condensates which contribute in the chiral
limit.  Finally, we study the effect of the topological charge $\Pi_{qg}(0)$ on the extraction of the meson mass and coupling from the lowest ${\cal L}_0^c$ moment.  

\d Comparing our mass prediction with the PDG Table quoted in Ref.\,\cite{PDG}, we expect that the $f_2(1950)$ or/and the $f'_2(2030)$ can have a sizeable hybrid $\bar qqg$ component in their wave functions. With the relative small value of the coupling $f_{2^+}$ compared to $f_\pi$, one may also use a Goldberg-Treiman-like relation to expect that the $2^{++}$  hadronic width into meson pairs of the $2^{++}$ hybrid state is large.  This feature may favour a large hybrid component of the $f_2(1950)$ or/and the $f'_2(2030)$ which have large hadronic widths.

\d We provide in Eq.\,\ref{eq:pi02} a first determination of the $2^{++}$ tensor hybrid topological charge at NLO. This value  could be checked from lattice QCD  or some low energy theorems (LET). \\
\section{{Acknowledgments}}
S. Narison wishes to thank Prof. V.I. Zakharov for a stimulating correspondence.
T.G. Steele is grateful for research funding from the Natural Sciences \& Engineering Research Council of Canada (NSERC Grant SAPIN-2021-00024).  

\vfill\eject
\input{bib_tensor26.tex}

\end{document}